\documentstyle[12pt]{article}

\topmargin -\headheight

\begin{document}

\title{Ground State of the Quantum Symmetric\\ Finite Size
XXZ Spin Chain\\ with Anisotropy Parameter $\Delta = \frac{1}{2}$}
\author{ V.~Fridkin$^{\rm a}$,  Yu.~Stroganov$^{\rm a,b}$ and
 D.~Zagier$^{\rm c}$\\\\
{$^{\rm a}$ Research Institute for Mathematical Sciences}\\
{ Kyoto University, Kyoto 606, Japan}\\
{$^{\rm b}$ Institute for High Energy Physics}\\
{ Protvino, Moscow region, Russia}\\
{$^{\rm c}$ Max-Planck-Institut fur Mathematik}\\
{ Gottfried-Claren-Strasse 26, D-53225, Bonn, Germany}}

\maketitle

\begin{abstract}
We find an  analytic solution of the Bethe Ansatz equations (BAE) for the
special case of a finite XXZ spin chain with free boundary conditions
and with a complex surface field which provides for $U_q(sl(2))$ symmetry
of the Hamiltonian. 
More precisely, we find one nontrivial solution, corresponding to 
the ground state of the system with anisotropy parameter $\Delta =
\frac{1}{2}$ corresponding to $q^3 = -1$.
\end{abstract}

\hfill \it Dedicated to Rodney Baxter \\
\hspace*{\fill}  on the occasion of his 60th birhday.
		
\vspace{0.5cm}

\rm It is widely accepted that the Bethe Ansatz equations for an  integrable
quantum spin chain can be solved analytically only in the thermodynamic
limit or for a small number of spin waves or short chains.
In this letter, however, we have managed to find a special solution of
 the BAE for a  spin chain of arbitrary length N with  N/2 spin waves.

It is well known (see, for example~\cite{ABBBQ} and references therein)
that there is a correspondence between the Q-state Potts Models and
the Ice-Type Models with anisotropy parameter $\Delta = \frac{\sqrt Q}{2}$.
The coincidence in the spectrum of an N-site self-dual Q-state quantum 
Potts chain with free ends with a part of the spectrum of the 
$U_q(sl(2))$ symmetrical 2N-site XXZ Hamiltonian (\ref{eq:b})
is to some extent a manifestation of this correspondence.
\begin{equation}
   \label{eq:b}
   H_{xxz}=\sum_{n=1}^{N-1}\{\sigma_n^{+}\sigma_{n+1}^{-}
   +\sigma_n^{-}\sigma_{n+1}^{+}
   +\frac{q+q^{-1}}{4}\,\sigma_n^z\sigma_{n+1}^z\
   +\frac{q-q^{-1}}{4}\,(\sigma_n^z-\sigma_{n+1}^z)\},
\end{equation}
where $\Delta=(q+q^{-1})/2$.
This Hamiltonian was considered by  Alcaraz {\it et al.}~\cite{ABBBQ}
and its $U_q(sl(2))$ symmetry was described  by Pasqier and 
Saleur~\cite{PS}.
The family of commuting transfer-matrices that commute with 
 $H_{xxz}$  was constructed by 
Sklyanin~\cite{SKL} incorporating a  method of Cherednik~\cite{CHER}. 

Baxter's T-Q equation for the case under consideration can be written 
as~\cite{ZH}  
\begin{equation}
 \label{eq:g}
 t(u)Q(u)=\phi(u+\eta/2)Q(u-\eta) + \phi(u-\eta/2)Q(u+\eta)
\end{equation}  
where $q=\exp{i\eta}$, $\phi(u)=\sin 2u\,\sin^{2N}u$ and $t(u)=\sin 2u\,T(u)$.
The $Q(u)$  are  eigenvalues of Baxter's auxilary matrix $\hat Q(u)$,
where $\hat Q(u)$ commutes with the transfer matrix $\hat T(u)$.
The eigenvalue $Q(u)$ corresponding to an eigenvector with $M=N/2-S_z$
 reversed spins has the form
\begin{equation}
 Q(u)= \prod\limits_{m=1}^M \sin (u-u_m) \sin (u+u_m).\nonumber
\end{equation}  
Equation (\ref{eq:g}) is equivalent to the Bethe Ansatz equations~\cite{MN}
\begin{equation}
\label{eq:i}
\left[\frac{\sin (u_k+\eta/2)}{\sin (u_k-\eta/2)}\right]^{2N}=
\prod\limits_{m\not=k}^M \frac{\sin (u_k-u_m+\eta)\sin (u_k+u_m+\eta)}
   {\sin (u_k-u_m-\eta)\sin (u_k+u_m-\eta)}.
\end{equation}  

In a recent article~\cite{BS} it was argued that the criteria for the above 
mentioned correspondence is the existence of a  second trigonometric 
solution for Baxter's T-Q equation and it was shown that in the case
$\eta=\pi /4$ the spectrum of $H_{xxz}$  contains the spectrum 
of the Ising model.
In this article we limit ourselves to the case $\eta = \pi /3$.
This case is in some sense  trivial since for this value of $\eta$, 
 $H_{xxz}$  corresponds to the 1-state Potts Model. 
We find  only one eigenvalue $T_0(u)$ of the transfer-matrices 
$\hat T(u)$ when Baxter's equation (\ref{eq:g}) has two independent
trigonometric solutions.
Solving for $ T(u)=T_0(u)$ analytically we find 
a trigonometric polynomial $Q_0(u)$ the zeros of which satisfy the
 Bethe Ansatz
equations (\ref{eq:i}). The number of spin waves is equal to $M=N/2$.
The corresponding eigenstate is the groundstate of $H_{xxz}$ with
 eigenvalue $E_0 = \frac{3}{2}(1-N)$, as discovered by Alcaraz 
{\it et al.}~\cite{ABBBQ} numerically.   

When does a second independent periodic solution 
exist?	
This question  was considered in article~\cite{BS}. Here we use a variation
more convenient for our goal.

Let us consider T-Q equation (\ref{eq:g}) for $\eta = \frac{\pi}{L}$, where
$L \ge 3$ is an integer.
Let us fix a sequence of spectral parameter values $v_k = v_0 + \eta k$,
where k are integers and write
 $\phi_k=\phi(v_k-\eta/2)$, $Q_k=Q(v_k)$ and $t_k=t(v_k)$.
The functions $\phi(u)$, $Q(u)$ and $t(u)$ are periodic with  period  $\pi$.
 So the sequences we have introduced are also periodic with  
period $L$, i.e.,  $\phi_{k+L}=\phi_k$, etc..

Setting $u=v_k$ in (\ref{eq:g}) gives the linear system
\begin{equation}
\label{eq:gd}
 t_k Q_k=\phi_{k+1} Q_{k-1} + \phi_{k} Q_{k+1}.
\end{equation} 
 The matrix of coefficients for this system has a tridiagonal form.
Taking $v_0 \ne \frac{\pi m}{2}$, where $m$ is an integer, we have 
$\phi_k \ne 0$ for all $k$.

It is straightforward to calculate the determinant of the 
$L-2 \times L-2$ minor obtained  by  deleting the two left most columns
 and two lower most rows. It is equal to the product 
$-\phi_1^2 \phi_2 \phi_3 \dots \phi_{L-1}$, which is nonzero,
hence  the rank of  $M$ cannot be less than $L-2$.
Here we are interested in the case when the rank of M is precisely $L-2$ and
 we have two linearly independent solutions for  equation (\ref{eq:gd}). 
Let us  consider  the three simplest cases L = 3, 4 and 5.
The parameter $\eta$ is equal to $\frac{\pi}{3}$, $\frac{\pi}{4}$
and $\frac{\pi}{5}$ respectively.

For $L=3$ the rank of $M$ is unity  and we immediately get
$t_0=-\phi_2,\> t_1=-\phi_0$ and $ t_2=-\phi_1$.
Returning to the functional form, we can write
\begin{equation}
\label{eq:t}
T_0(u)=t_0(u)/\sin 2u=-\phi (u+\pi/2)/\sin 2u = \cos^{2N} u.
\end{equation} 
 This is the unique eigenvalue of the transfer-matrix for which the 
T-Q equation has two independent periodic solutions.
  It is well known (see, for example,~\cite{MN}) that the eigenvalues of 
$H_{xxz}$ are related to the eigenvalues $t(u)$ by
\begin{equation}
E=-\cos \eta (N+2-\tan^2 \eta) + \sin \eta 
\frac{t^{\prime}(\eta /2)}{t(\eta /2)}.\nonumber
\end{equation} 
For the eigenstate corresponding to eigenvalue (\ref{eq:t}) we obtain 
$E_0=3/2(1-N)$.
This is the groundstate energy which was
 discovered  by Alcaraz {\it et al.}~\cite{ABBBQ} numerically.   

Below we find all solutions of Baxter`s T-Q equation 
corresponding to $T(u)=T_0(u)$.
Zeros of these solutions satisfy the BAE (\ref{eq:i}). 
In particular we find  $Q_0(x)$ corresponding to physical 
Bethe state.

For $L=4$,
deleting the second row and the forth column of $M$ we obtain a minor with 
determinant $-\phi_0 \phi_3 (t_0+t_2)$. It is zero when $t_2=-t_0$, i.e.,
 $t(u+\frac{\pi}{2})=-t(u)$.
Considering the other minors we obtain the functional equation
\begin{equation}
t(u+\pi/8)t(u-\pi/8)=\phi(u+\pi/4)\phi(u-\pi/4)-
\phi(u)\phi(u+\pi/2). \nonumber  
\end{equation}
This functional equation was used in~\cite{BS} to find  $t(u)$
and  show that this part of the spectrum of $H_{xxz}$ coincides with the
 Ising model.
It would be interesting to find a corresponding $Q(u)$.

Lastly for $L=5$, minor $M_{35}$ (the third row and the fifth column are 
deleted) has determinant 
$\phi_0 \phi_4 (t_0 t_1 + \phi_1 t_3 - \phi_0 \phi_2)$.
Setting this to zero we have
\begin{equation}
t(u) t(u+\pi/5) + \phi(u+\pi/10) t(u+3\pi/5)
 - \phi(u-\pi/10) \phi(u+3 \pi/10)=0.\nonumber
\end{equation}
It is not difficult to check that in this case all  $4 \times 4$ minors have
zero determinant and that the rank of M is 3. Thus we have two independent
periodic solutions of Baxter's T-Q equation.

Note that this functional relation coincides with the 
Baxter-Pearce relation for the  hard hexagon model~\cite{BP82}.
We have obtained the same truncated functional relations that
have been obtained in \cite{BS} with the same assumptions.

We now consider the solution of Baxter's Equation for $\eta = \frac {\pi}{3}$
and $T=T_0$.	 
For  $\eta = \frac{\pi}{3}$ and transfer-matrix eigenvalue 
 $T_0(u)=\cos^{2N} u$, the T-Q equation (\ref{eq:g}) reduces to 
\begin{equation}
\label{eq:phiQ}
\phi(u+3\eta/2) Q(u) + \phi(u-\eta/2) Q(u+\eta) + \phi(u+\eta/2) Q(u-\eta)=0.
\end{equation} 
This equation  can be rewritten as
\begin{equation}
\label{eq:no3}
f(v)+f(v+2\pi/3)+f(v+4\pi/3) = 0,
\end{equation}
where $f(v) = \sin v\> \cos^{2N} (v/2)\> Q(v/2)$ has period $2\pi$.
The trigonometric polynomial $f(v)$ is an odd function, so it can be written
\begin{equation}
\label{eq:ff}
f(v) = \sum_{k=1}^{K} c_k \sin kv,
\end{equation}
where $K$ is the degree of $f(v)$.
Then equation (\ref{eq:no3}) is equivalent to $c_{3m}=0,\> m \in  Z$.

The condition that $f(v)$ be divisible by $\sin v \cos^{2N}(v/2)$ is
equivalent to 
\begin{equation}
(\frac{d}{dv})^i f(v)|_{v=\pi} = 0,\qquad i = 0,1,\dots ,2N.\nonumber
\end{equation}
For even $i$ this condition is immediate, whereas for $i=2j-1$ we  use  
(\ref{eq:ff}) to obtain
\begin{equation}
\label{eq:ls}
\sum_{k=1, k \ne 3m}^{K} (-1)^k c_k k^{2j-1} = 0,\qquad j=1,2,\dots ,N.
\end{equation}

Our problem is thus to find $\{c_k\}$ satisfying the last equation.
This problem is a special case of a more general problem which can be
 formulated as follows.
Given a set of different complex numbers $X=\{x_1,x_2,\dots,x_I\}$
we seek another complex set $B=\{\beta_1,\beta_2,\dots,\beta_I\}$
where $\beta_i \ne 0$ for some $i$, so that 
\begin{equation}
\label{eq:prob}
\sum_{i=1}^{I} \beta _i P(x_i) = 0
\end{equation}
for any polynomial $P(x)$ of degree not more than $N-1$. 
It is clear that for $I \le N$ the system $B$ does not exist.
If $\beta_1 \ne 0$, for example, the product 
$(x-x_2)(x-x_3)\dots(x-x_I)$ provides a counterexample.

Let $I=N+1$. We try the polynomials
\begin{equation}
\label{eq:pol}
P_{r} = \prod_{i=1, i \ne r,}^{N} (x-x_i), \qquad r=1,2,\dots, N.
\end{equation}  
Condition (\ref{eq:prob}) gives
$\beta_r P_{r}(x_r) + \beta_I P_{r}(x_I) = 0$
and we immediately obtain 
\begin{equation}
\label{eq:sol}
\beta_r = \mbox{const} \prod_{i=1, i \ne r}^{N+1}(x_r-x_i)^{-1},
\end{equation}
which is a solution  because  the system (\ref{eq:pol}) forms 
a basis of the linear space of $N-1$ degree polynomials. 
So for $I=N+1$ we have a unique solution
 (up to an arbitrary nonzero constant) given by (\ref{eq:sol}).
It is easy to show that for $I=N+\nu$ we obtain a $\nu$-dimensional
linear space of solutions.
 
Returning to (\ref{eq:ls}) we  consider $N=2n$,
$n$ a positive integer. Fix $I=N+1=2n+1$. The degree $K$ becomes $3n+1$. 
It is convenient to use a new index
$k=|3\kappa+1|$ , where $|\kappa| \le n$. 
Equation (\ref{eq:ls}) can be rewritten as 
\begin{equation}
\sum_{\kappa=-n}^{n} \beta_{\kappa}(3\kappa+1)^{2(j-1)}=0,
 \qquad j=1,2,\dots,N,\nonumber
\end{equation}
where we use new unknowns
$\beta_{\kappa}=(-1)^{\kappa} c_{|3\kappa +1|} |3\kappa+1|$
instead of $c_k$.
Using  (\ref{eq:sol}) and (\ref{eq:ff}) we obtain the function $f(v)$
\begin{equation}
\label{eq:sol5}
f(v)  = \sum_{\kappa = -n}^{n}(-1)^{\kappa}
\biggl(\begin{array}{c}
2n + \frac{2}{3} \\
n - \kappa
\end{array}\biggr)\biggl(\begin{array}{c}
2n - \frac{2}{3} \\
n + \kappa
\end{array}\biggr) \sin (3\kappa + 1)v.
\end{equation}

We recall that the solution of Baxter's T-Q equation for $T(u)=T_0(u)$
 is given by
\begin{equation} 
Q_0 (u) = f(2u)/(\sin 2u \cos^{2N} u)
\end{equation}
and its zeros $\{u_k\}$ satisfy the BAE (\ref{eq:i}).  

Another way to derive the above solution is to observe that the function 
$f(v)$ satisfies a simple second order linear differential equation.
    Indeed, it is easily seen that the functions $F^+(x)$ and $F^-(x)$, where
\begin{equation}
F^+(x)  = \sum_{\kappa = -n}^{n}(-1)^{\kappa}
\biggl(\begin{array}{c}
2n + \frac{2}{3} \\
n - \kappa
\end{array}\biggr)\biggl(\begin{array}{c}
2n - \frac{2}{3} \\
n + \kappa
\end{array}\biggr) x^{\kappa + \frac{1}{3}} \mbox{ and }  F^-(x)=F^+(1/x).
\end{equation}
are the two linearly independent solutions of the differential equation
\begin{equation}
\label{eq:diff} 
\{((\theta + n)^2 - 1/9)/x + 
(\theta - n)^2 - 1/9
\} F^{+}=0,
\end{equation}
where $\theta=x\frac{d}{dx}$.\footnote{Up to a change of variables this is just the standard
    hypergeometric differential equation, and in fact 
$F^+(x)= \mbox{const } F(-2n,2/3-2n,5/3,-x) x^{1/3-n} $ }
  Now the fact that there is 
a combination $f(v)$ of $F^+(e^{3iv})$
    and $F^-(e^{3iv})$ which vanishes to order $2N+1$ at $v=\pi$ follows
 immediately
    from the fact that $x=-1$ is a singular point of the differential equation 
    (\ref{eq:diff})
 and that the indicial equation at this point has roots 0 and $2n+1$. 
      In terms of the variable $v$,  equation (\ref{eq:diff}) becomes 
\begin{equation}
\label{eq:diftb} 
\frac{d^2f}{dv^2} + 6n\tan (3v/2)\frac{df}{dv} + (1 - 9 n^2)\>f=0.
\end{equation}

The zeros of $f(v)$, the density of which is important in the thermodynamic
limit, are located on the imaginary axis in the complex $v$-plane.
So it is convenient to make the change of variable $v=i s$.
It is also useful to introduce another function 
$g(s) = f(is)/\cosh^{2n}(3s/2)$.
The differential equation for $g(s)$ is then
\begin{equation}
\label{eq:diftbb} 
g^{\prime \prime} + \left(\frac{9n(2n+1)}{2\cosh^2(3s/2)}-1\right)g = 0.
\end{equation}
Let $g(s_0)=0$. For large $n$ we have in a small vicinity of $s_0$
an approximate equation $g^{\prime \prime} + \omega_0^2 g = 0$.
This equation describes a  harmonic oscillator with  frequency 
$\omega_0=3n/\cosh(3s_0/2)$.
The  distance between nearest zeros is approximately 
$\Delta s=\pi/\omega$ and we obtain the following density function
 which describes the  number of zeros per unit length 
\begin{equation}
\label{eq:den} 
\rho(s)=1/\Delta s=\omega/\pi=3n/(\pi \cosh(3s/2)).
\end{equation}

 We note that equation  (\ref{eq:diftbb}) has a history as rich as the BAE.
Eckart \cite{Eck30} used the Schrodinger equation
 with bell-shaped potential $V(r) = -G/\cosh^{2}r$ for phenomenological
 studies in atomic and molecular physics. Later it was used in chemistry,
biophysics and  astrophysics, just to name a few. For more recent references
 see, for example, \cite{Zn}. 
 
We are grateful to  V.~Bazhanov, A.~Belavin, L.~Faddeev and G.~Pronko
for useful discussions.
We would like to thank M.~Kashiwara and T.~Miwa for their kind hospitality
 in RIMS.
This work is supported in part by RBRF--98--01--00070,
INTAS--96--690 (Yu.~S.). V.~F. is supported by a JSPS fellowship.

\end{document}